\documentclass[zpreprint,zbstdefault]{zeus_paper}
\usepackage[english]{babel}
\newcommand{\ZcoosysA}{%
The ZEUS coordinate system is a right-handed Cartesian system, with the $Z$
axis pointing in the proton beam direction, referred to as the ``forward
direction'', and the $X$ axis pointing left towards the center of HERA.
The coordinate origin is at the nominal interaction point.\xspace}

\newcommand{\ZcoosysfnA}{\footnote{\ZcoosysA}}

\newcommand{\Zdetdesc}{%
A detailed description of the ZEUS detector can be found 
elsewhere~\cite{zeus:1993:bluebook}. A brief outline of the 
components that are most relevant for this analysis is given
below.\xspace}

\newcommand{\Zctdmalc}[1]{%
Charged particles are tracked in the central tracking detector (CTD)~\citeCTD,
which operates in a magnetic field of $1.43\Tesla$ provided by a thin 
superconducting solenoid. The central tracking detector consists of 72~cylindrical drift chamber 
layers, organized in nine superlayers covering the polar-angle#1 region 
\mbox{$15^\circ<\theta<164^\circ$}. The transverse-momentum resolution for
full-length tracks is $\sigma(p_T)/p_T=0.0058p_T\oplus0.0065\oplus0.0014/p_T$,
with $p_T$ in $\Gev$.}

\newcommand{\Zcalmalc}{%
The high-resolution uranium--scintillator calorimeter (CAL)~\citeCAL consists 
of three parts: the forward, the barrel and the rear
calorimeters. Each part is subdivided transversely into towers and
longitudinally into one electromagnetic section and either one (in rear)
or two (in barrel and forward) hadronic sections. The smallest subdivision of
the calorimeter is called a cell.  The calorimeter energy resolutions, as measured under
test-beam conditions, are $\sigma(E)/E=0.18/\sqrt{E}$ for electrons and
$\sigma(E)/E=0.35/\sqrt{E}$ for hadrons, with $E$ in $\Gev$.}

\chardef\usc=95
\chardef\til=126
\catcode`\@=11 
\DeclareRobustCommand\xdotspace{\futurelet\@let@token\@xdotspace}
\def\@xdotspace{%
  \ifx\@let@token.\else
  \ifx\@let@token\bgroup.\else
  \ifx\@let@token\egroup.\else
  \ifx\@let@token\/.\else
  \ifx\@let@token\ .\else
  \ifx\@let@token~.\else
  \ifx\@let@token!.\else
  \ifx\@let@token,.\else
  \ifx\@let@token:.\else
  \ifx\@let@token;.\else
  \ifx\@let@token?.\else
  \ifx\@let@token/.\else
  \ifx\@let@token'.\else
  \ifx\@let@token).\else
  \ifx\@let@token-.\else
  \ifx\@let@token\@xobeysp.\else
  \ifx\@let@token\space.\else
  \ifx\@let@token\@sptoken.\else
   .\space
   \fi\fi\fi\fi\fi\fi\fi\fi\fi\fi\fi\fi\fi\fi\fi\fi\fi\fi}
\catcode`\@=12 

\newcommand{\stru}[2]{%
   \relax\ifmmode\hbox{\vrule height#1 depth#2 width0pt}%
   \else\vrule height#1 depth#2 width0pt\fi}

\newcommand{\Ronum}[1]{\uppercase\expandafter{\romannumeral#1}}
\newcommand{\ronum}[1]{\expandafter{\romannumeral#1}}
\DeclareRobustCommand{\LaTeXZ}{%
  \LaTeX\kern-.05em4\kern-.1em
  {\raisebox{-0.2ex}{$\scriptstyle\text{ZEUS}$}}\xspace}

\DeclareMathAlphabet{\mathbf}{OT1}{cmr}{bx}{sl}
\newcommand{\eVdist}{\kern-0.06667em}

\newcommand{\Gev}{{\text{Ge}\eVdist\text{V\/}}}

\newcommand{\gev}{{\,\text{Ge}\eVdist\text{V\/}}}

\newcommand{\pbi}{\,\text{pb}^{-1}}

\newcommand{\Tesla}{\,\text{T}}

\newcommand{\slashfrac}[2]{%
  \raisebox{0.5ex}{\ensuremath #1}\kern-0.12em/\kern-0.08em
  \raisebox{-.8ex}{\ensuremath #2}}

\newcommand{\sqr}[3]{%
    {\vcenter{\hrule height.#3ex\hbox{\vrule width.#2ex height#1ex
     \kern#1ex\vrule width.#3ex}\hrule height.#2ex}}}

\catcode`\@=11 
\newcommand{\parenbar}{\mathpalette\p@renb@r}
\def\p@renb@r#1#2{\vbox{%
  \ifx#1\scriptscriptstyle \dimen@.7em\dimen@ii.2em\else
  \ifx#1\scriptstyle \dimen@.8em\dimen@ii.25em\else
  \dimen@1em\dimen@ii.4em\fi\fi \offinterlineskip
  \ialign{\hfill##\hfill\cr
    \vbox{\hrule width\dimen@ii}\cr
    \noalign{\vskip-.3ex}%
    \hbox to\dimen@{$\mathchar300\hfil\mathchar301$}\cr
    \noalign{\vskip-.3ex}%
    $#1#2$\cr}}}
\catcode`\@=12 

\newcommand{\diff}{{\rm d}}

\newcommand{\IP}{{\rm I$\kern-0.01667em$P}\xspace}

\mathchardef\qsm=63
\mathchardef\pls=43
\mathchardef\mns=512
\mathchardef\plm=518
\mathchardef\eql=61
\mathchardef\smallleft=300
\mathchardef\smallright=301
\mathchardef\les=316
\mathchardef\gre=318
\mathchardef\leq=532
\mathchardef\grq=533

\catcode`\@=11 
\newcounter{pict@width}
\newcounter{pict@height}
\newlength{\pict@scale}
\newcommand{\psfigadd}[4]{%
\setcounter{pict@width}{1*\ratio{#2+\pict@scale/2}{\pict@scale}}
\setcounter{pict@height}{1*\ratio{#3+\pict@scale/2}{\pict@scale}}
\setlength{\unitlength}{\pict@scale}
\hbox to #2{\hspace{-\fill}\begin{picture}(\thepict@width,\thepict@height)
\put(0,0){\psfig{figure=#1,width=#2,height=#3,clip=}}
\SetScale{0.283466457}
\SetWidth{1.763889}
{#4}
\end{picture}}
}
\newcounter{pict@widthfst}
\newcounter{pict@widthscd}
\newcounter{pict@widthtot}
\newcommand{\psfigaddtwo}[7]{%
\setcounter{pict@widthfst}{1*\ratio{#2+\pict@scale/2}{\pict@scale}}
\setcounter{pict@widthscd}{1*\ratio{#2+#4+\pict@scale/2}{\pict@scale}}
\setcounter{pict@widthtot}{1*\ratio{#2+#4+#6+\pict@scale/2}{\pict@scale}}
\setcounter{pict@height}{1*\ratio{#3+\pict@scale/2}{\pict@scale}}
\setlength{\unitlength}{\pict@scale}
\hbox{\hspace{-\fill}\begin{picture}(\thepict@widthtot,\thepict@height)
\put(0,0){\psfig{figure=#1,width=#2,height=#3,clip=}}
\put(\thepict@widthscd,0){\psfig{figure=#5,width=#6,height=#3,clip=}}
\SetScale{0.283466457}
\SetWidth{1.763889}
{#7}
\end{picture}}
}
\newcommand{\psfigror}[4]{%
\setcounter{pict@width}{1*\ratio{#2+\pict@scale/2}{\pict@scale}}
\setcounter{pict@height}{1*\ratio{#3+\pict@scale/2}{\pict@scale}}
\setlength{\unitlength}{\pict@scale}
\hbox{\begin{picture}(\thepict@width,\thepict@height)
\put(0,\thepict@height){\psfig{figure=#1,width=#3,height=#2,clip=,angle=270}}
\SetScale{0.283466457}
\SetWidth{1.763889}
{#4}
\end{picture}}
}
\newcommand{\psfigrol}[4]{%
\setcounter{pict@width}{1*\ratio{#2+\pict@scale/2}{\pict@scale}}
\setcounter{pict@height}{1*\ratio{#3+\pict@scale/2}{\pict@scale}}
\setlength{\unitlength}{\pict@scale}
\hbox{\begin{picture}(\thepict@width,\thepict@height)
\put(0,0){\psfig{figure=#1,width=#3,height=#2,clip=,angle=90}}
\SetScale{0.283466457}
\SetWidth{1.763889}
{#4}
\end{picture}}
}
\catcode`\@=12 
\newlength\listtextwidth

\catcode`\@=11 
\newlength{\@tabfninsert}
\newlength{\@tabfnwidth}
\newcommand{\tabfootnote}[2]{%
  \setlength{\@tabfninsert}{0.8em}
  \setlength{\@tabfnwidth}{\textwidth}
  \addtolength{\@tabfnwidth}{-\@tabfninsert}
  \addtolength{\@tabfnwidth}{-0.4em}
  \noindent\makebox[\@tabfninsert][r]{\footnotesize$^{#1}$\hfil}\hfill%
  \parbox[t]{\@tabfnwidth}{\footnotesize #2\hfill}}
\catcode`\@=12 

\newcommand {\dstar} {D^{\ast \pm}}

\def\citeCTD{{\cite{%
nim:a279:290,*npps:b32:181,*nim:a338:254%
}}\xspace}
\def\citeCAL{{\cite{%
nim:a309:77,*nim:a309:101,*nim:a321:356,*nim:a336:23%
}}\xspace}

\begin{document}
\prepnum{DESY-03-221}

\title{
Photoproduction of $\mathbf{\dstar}$ mesons \\
associated with a leading neutron \\
}                                                       
                    
\author{ZEUS Collaboration}
\date{\today}

\abstract{
The photoproduction of $\dstar (2010)$ mesons 
associated with a leading neutron has been observed 
with the ZEUS detector
in $ep$ collisions at HERA using an integrated luminosity of
$80$~pb$^{-1}$.
The neutron carries a large fraction, \mbox{$x_L>0.2$}, of the incoming
proton beam energy and is detected at very
small production angles, \mbox{$\theta_n<0.8$ mrad}, an indication of
peripheral scattering. The $D^*$ meson 
is centrally produced with 
pseudorapidity \mbox{$|\eta|<1.5$}, and has 
a transverse momentum \mbox{$p_{\it T} > 1.9$ GeV}, which is large compared to
the average transverse momentum of the neutron of 0.22 GeV.
The ratio of neutron-tagged to inclusive $D^*$ production is
 \mbox{$8.85\pm 0.93({\rm stat.})^{+0.48}_{-0.61}({\rm syst.})$ \%} 
in the photon-proton center-of-mass energy range 
\mbox{$130 <W<280$ GeV}. The data suggest that the presence of a 
hard scale 
enhances the fraction of events with a leading neutron 
in the final state.
}

\makezeustitle

\def\3{\ss}                                                                                        
\pagenumbering{Roman}                                                                              
                                                   %
\begin{center}                                                                                     
{                      \Large  The ZEUS Collaboration              }                               
\end{center}                                                                                       
  S.~Chekanov,                                                                                     
  M.~Derrick,                                                                                      
  D.~Krakauer,                                                                                     
  J.H.~Loizides$^{   1}$,                                                                          
  S.~Magill,                                                                                       
  S.~Miglioranzi$^{   1}$,                                                                         
  B.~Musgrave,                                                                                     
  J.~Repond,                                                                                       
  R.~Yoshida\\                                                                                     
 {\it Argonne National Laboratory, Argonne, Illinois 60439-4815}, USA~$^{n}$                       
\par \filbreak                                                                                     
  M.C.K.~Mattingly \\                                                                              
 {\it Andrews University, Berrien Springs, Michigan 49104-0380}, USA
% ...
\newpage                                                                                           
$^{\    1}$ also affiliated with University College London, London, UK \\                          
$^{\    2}$ retired \\                                                                             
% ...
$^{\    5}$ now at Dongshin University, Naju, Korea \\                                             
% ...
$^{\    8}$ partly supp. by the Israel Sci. Found. and Min. of Sci.,                               
and Polish Min. of Scient. Res. and Inform. Techn., grant no.2P03B12625\\                          
% ...
                                                           %
                                                           %
\newpage   
                                                           %
                                                           %
\begin{tabular}[h]{rp{14cm}}                                                                       
$^{a}$ &  supported by the Natural Sciences and Engineering Research                               
          Council of Canada (NSERC) \\                                                             
$^{b}$ &  supported by the German Federal Ministry for Education and                               
          Research (BMBF), under contract numbers HZ1GUA 2, HZ1GUB 0, HZ1PDA 5, HZ1VFA 5\\         
$^{c}$ &  supported by the MINERVA Gesellschaft f\"ur Forschung GmbH, the                          
          Israel Science Foundation, the U.S.-Israel Binational Science                            
          Foundation and the Benozyio Center                                                       
          for High Energy Physics\\                                                                
$^{d}$ &  supported by the German-Israeli Foundation and the Israel Science                        
          Foundation\\                                                                             
$^{e}$ &  supported by the Italian National Institute for Nuclear Physics (INFN) \\                
$^{f}$ &  supported by the Japanese Ministry of Education, Culture,                                
          Sports, Science and Technology (MEXT) and its grants for                                 
          Scientific Research\\                                                                    
$^{g}$ &  supported by the Korean Ministry of Education and Korea Science                          
          and Engineering Foundation\\                                                             
$^{h}$ &  supported by the Netherlands Foundation for Research on Matter (FOM)\\                   
$^{i}$ &  supported by the Polish State Committee for Scientific Research,                         
          grant no. 620/E-77/SPB/DESY/P-03/DZ 117/2003-2005\\                                      
$^{j}$ &  partially supported by the German Federal Ministry for Education                         
          and Research (BMBF)\\                                                                    
$^{k}$ &  partly supported by the Russian Ministry of Industry, Science                            
          and Technology through its grant for Scientific Research on High                         
          Energy Physics\\                                                                         
$^{l}$ &  supported by the Spanish Ministry of Education and Science                               
          through funds provided by CICYT\\                                                        
$^{m}$ &  supported by the Particle Physics and Astronomy Research Council, UK\\                   
$^{n}$ &  supported by the US Department of Energy\\                                               
$^{o}$ &  supported by the US National Science Foundation\\                                        
$^{p}$ &  supported by the Polish State Committee for Scientific Research,                         
          grant no. 112/E-356/SPUB/DESY/P-03/DZ 116/2003-2005,2 P03B 13922\\                       
$^{q}$ &  supported by the Polish State Committee for Scientific Research,                         
          grant no. 115/E-343/SPUB-M/DESY/P-03/DZ 121/2001-2002, 2 P03B 07022\\                    
\end{tabular}                                                                                      
                                                           %
                                                           %

\newcommand{\lap}{\ensuremath{\stackrel{_{\scriptstyle <}}{_{\scriptstyle\sim}}}}
\newcommand{\gap}{\ensuremath{\stackrel{_{\scriptstyle >}}{_{\scriptstyle\sim}}}}
\newcommand{\Journal}[4]{{#1}{\bf #2}, #4 (#3)}

\newcommand{\CPC}{Comp.\ Phys.\ Comm.\ }
\newcommand{\EPA}{Eur.\ Phys.\ J.\ {\bf A}}
\newcommand{\EPC}{Eur.\ Phys.\ J.\ {\bf C}}
\newcommand{\MPL}{Mod.\ Phys.\ Lett.\ }
\newcommand{\NCA}{Nuovo Cimento A}
\newcommand{\NIM}{Nucl.\ Instr.\ and Meth.\ }
\newcommand{\NIMA}{Nucl.\ Instr.\ and Meth.\ {\bf A}}
\newcommand{\NP}{Nucl.\ Phys.\ }
\newcommand{\NPA}{Nucl.\ Phys.\ {\bf A}}
\newcommand{\NPB}{Nucl.\ Phys.\ {\bf B}}
\newcommand{\NPPS}{Nucl.\ Phys.\ Proc.\ Suppl.\ }
\newcommand{\PiNL}{$\Pi$N Newslett.\ }
\newcommand{\PHM}{Phil.\ Mag.\ }
\newcommand{\PL}{Phys.\ Lett.\ }
\newcommand{\PLB}{Phys.\ Lett.\ {\bf B}}
\newcommand{\PR}{Phys.\ Rev.\ }
\newcommand{\PRep}{Phys.\ Rep.\ }
\newcommand{\PRL}{Phys.\ Rev.\ Lett.\ }
\newcommand{\PRP}{Phys.\ Rep.\ }
\newcommand{\PRC}{Phys.\ Rev.\ {\bf C}}
\newcommand{\PRD}{Phys.\ Rev.\ {\bf D}}
\newcommand{\ProgPNP}{Prog.\ Part.\ Nucl.\ Phys.\ }
\newcommand{\RMP}{Rev.\ Mod.\ Phys.\ }
\newcommand{\SNP}{Sov.\ J.\ Nucl.\ Phys.\ }
\newcommand{\ZPA}{Z.\ Phys.\ {\bf A}}
\newcommand{\ZPC}{Z.\ Phys.\ {\bf C}}

\pagenumbering{arabic} 
\pagestyle{plain}
\section{Introduction}
\label{sec-int}

Events containing
a  leading neutron have been studied 
in $ep$ collisions at HERA 
\cite{pl:b384:388,np:b596:3,np:b637:3,epj:c6:587}.
The neutrons carry a large fraction of the incoming proton beam energy,
$x_L>0.2$,
and are produced at very small scattering angles, $\theta_n<0.8$ mrad,
indicative of a peripheral process.


The small transverse momenta ($p_{\it T}$) which characterize
 leading baryon production processes 
imply a soft scale, which means
 that a non-perturbative approach is required to model such events.
Particle-exchange models within  Regge theory 
\cite{pdbcollins}, in particular the one-pion-exchange 
model (OPE) \cite{yukawa,sullivan,bishari},
are often applied 
to describe leading neutron production.
Charm production,  in contrast, can be used to investigate
parton dynamics 
because the  charm-quark mass provides the hard scale necessary to ensure
the applicability of perturbative Quantum Chromodynamics (pQCD). 
Therefore the study of charm production in events with a leading
neutron gives information on the
interplay between soft and hard scales.

This letter presents measurements of
$\dstar$ photoproduction associated with a leading neutron.
Differential cross sections and ratios to inclusive 
$\dstar$ photoproduction are reported.  These results extend
previous ZEUS studies \cite{pl:b384:388,np:b596:3,np:b637:3} of leading-neutron
production in dijet and inclusive photoproduction 
and deep inelastic scattering.



\section{Charm and neutron production}

An important process in charm photoproduction is  Boson-Gluon Fusion
(BGF). At leading  order this corresponds to the direct component
where the photon
couples  directly to a high-transverse-momentum $c\bar{c}$ pair which 
interacts with a gluon from the proton.
Another contribution to the cross section comes from the resolved
component, where the photon  acts as a source of partons, interacting
with the proton mostly via charm excitation processes, 
\mbox{$cq \rightarrow cq$}  and \mbox{$cg \rightarrow cg$} \cite{pl:b565:87}. 

The mechanism for leading neutron production is not 
well understood. In the following subsections some models for neutron 
production are briefly discussed.

\subsection{Fragmentation models}

In fragmentation models of partons into hadrons, such as the
cluster model \cite{cluster} or the Lund string 
model \cite{anderson},  a certain 
fraction of neutrons is expected 
in the final state.
In this case the leading neutrons are produced by fragmentation of the
proton remnant using the same 
mechanism as is used for the other final state hadrons. 
Such models predict a softer $x_L$ distribution than that measured 
 \cite{pl:b384:388,np:b596:3,np:b637:3,epj:c6:587}.

\subsection{One-pion-exchange model}
\label{sec:opeint}

Previous studies have shown that particle-exchange models 
\cite{yukawa,sullivan,bishari,field,ganguli,zakharov,zoller}
describe data on leading neutron production both 
at HERA\cite{pl:b384:388,np:b596:3,np:b637:3,epj:c6:587} and 
at hadroproduction experiments
~\cite{erwin,pickup,engler,robinson,flauger,hanlon,hartner,eisenberg,blobel,abramowicz}. 
In such models the transition amplitude for $p \rightarrow n$ is 
dominated by OPE  and the electroproduction cross section can be 
written as the convolution of a function 
describing the splitting of a proton into a $\pi n$ system, i.e.
the pion flux factor $f_{\pi /p}(x_L, t)$,
and the $e\pi$ cross section:
\begin{equation}
\frac{d\sigma_{ep \rightarrow e^\prime nX}}{dx_L\, dt} = 
f_{\pi /p}(x_L, t) \sigma^{e\pi}(s^\prime),
\label{eq:opef}
\end{equation}
where $t$ is the squared four-momentum transfer at the proton vertex, 
$s^\prime = s(1-x_L)$ is the squared 
 center-of-mass energy of the $e\pi$  system and
$s$ is that of the $ep$.

The flux factors found in the literature can be expressed
in general as~\cite{bishari,field} 
\begin{equation}
f_{\pi /p}(x_L,t)=  \frac{1}{4\pi} \frac{2g_{p\pi p}^2}{4\pi}
                         \frac{-t}{(t-m^2_{\pi})^2}
                          (1-x_L)^{1-2\alpha(t)}
                         \left[ F(x_L ,t) \right]^2,
\label{eq:flux}
\end{equation}
where $g_{p\pi p}^2/(4\pi) \sim 14.5$ is the $p \pi p$ coupling constant,
 $m_\pi$ is the pion mass and $\alpha(t)$ is defined below.
The form-factor $F(x_L ,t)$ accounts for the finite size of the 
nucleon and pion. 
Examples of  flux factor parametrizations are:

\begin{itemize}

\item{ $f_1$ \cite{holtmann}:}
\begin{equation}
F(x_L,t)=\exp\left[R^2 \frac{(t-m_\pi^2)}{(1-x_L)}\right],\ \alpha(t)=0,
\label{flu2}
\end{equation}
where $R=0.6$ GeV$^{-1}$ and $F(x_L,t)$ is the light-cone form factor.


\item{ $f_2$ \cite{bishari}:}
\begin{equation}
F(x_L,t)=1,\ \alpha(t) = \alpha_{\pi}(t)
\label{flu3}
\end{equation}
where $\alpha_{\pi}(t) \simeq t$  (with $t$ in GeV$^{2}$) is the Regge trajectory of the pion.


\item{$f_3$ \cite{kopeli}:}
\begin{equation}
F(x_L,t)=\exp[b(t-m_\pi^2)],\ \alpha(t)  = \alpha_{\pi}(t)
\label{fluc}
\end{equation}
where $b=0.3$ GeV$^{-2}$ and $F(x_L,t)$ is the exponential
form factor ;


\item{$f_4$ \cite{frank89}:}
\begin{equation}
F(x_L,t)=\frac{(1-m_\pi^2/\Lambda^2)}{(1-t^2/\Lambda^2)},\ \alpha(t)=0
\label{flu4}
\end{equation}
where $ \Lambda=0.25$ GeV and $F(x_L,t)$ is the monopole form factor.

\end{itemize}

The term $\sigma^{e\pi}$ in Eq.~(\ref{eq:opef})  involve
the parton distribution in the pion.
Charm production in association with a leading neutron
is potentially sensitive to the gluon content of the pion in OPE models
via the BGF process.
Parametrizations for the
pionic parton distribution function (PDF)  available 
in the literature were obtained by performing
fits to $\pi N$ scattering data, assuming some parametrization for the
nucleon structure function. Examples of such parametrizations 
 are those by Owens \cite{owens}
which come from fits on  $J/\Psi$ and dimuon production data. 
The more recent GRV parametrizations \cite{grv} assume a  valence-like
structure for the pion at a certain low scale. This distribution is
dynamically evolved and the results combined with the constraints imposed
 by prompt photon production data on the pionic gluon density.

\subsection{Rescattering effects}
\label{sec:resc}

From the factorization hypothesis expressed in  Eq.~(\ref{eq:opef}),
it is expected that the ratios, $r$, of neutron-tagged 
to inclusive cross sections for different electroproduction processes 
are about the same. Most of the dependence of the cross sections on the
kinematics of the processes  cancels;  remaining 
differences can be attributed to differences between the pion and
proton energies and their PDFs.
However, larger differences than these may arise from
neutron absorption, which can occur
through rescattering of the neutron on the exchanged
 photon\cite{zakharov,rescat}.
With increasing  size of the virtual photon 
more rescattering may be expected.

Inclusive photoproduction cross sections are well described by vector meson
dominance models\cite{bauer,np:b627:3,h1new}, where the dipole
associated with 
the photon  is of hadronic size. In dijet photoproduction,
the presence of the hard scale given by the transverse energy of the 
jets implies smaller dipole size. In the infinite-momentum frame
the smaller dipole size corresponds to the enhancement of the 
direct photon component
at high transverse energy $E_T$. 
Such an enhancement has been observed at HERA \cite{dij}.
Absorptive effects in dijet photoproduction, therefore, are 
expected to be smaller than those in inclusive photoproduction.
In the case of charm photoproduction, an additional hard scale 
is provided by the mass of the charm quark.  
Previous ZEUS measurements 
have shown that requiring the presence of charm further suppresses 
the resolved component 
\cite{epj:c6:67} compared to inclusive dijet photoproduction.
 Therefore rescattering in $D^*$ photoproduction may be further
suppressed in comparison  to dijet photoproduction.
In deep inelastic scattering (DIS), at sufficiently high photon
virtuality,
little rescattering should occur. 
Within such a picture, therefore, the ratios  are expected to have the following relationship:
\[
r^{\gamma p} < r^{\rm jj} < r^{\rm D^*} \lesssim {\rm r}^{\rm DIS}.
\]


\section{Experimental conditions}
\label{sec-exp}

The integrated luminosity of $80.2 \pm 1.8 \pbi$ used for this measurement was
collected at the $ep$ collider HERA with the ZEUS detector during 1998 - 
2000, when HERA collided $27.5\gev$  
 electrons or positrons\footnote{Hereafter, both $e^+$ and $e^-$
are referred to as electrons.}
 with $920\gev$ protons, giving a
center-of-mass energy of 318 GeV.

\Zdetdesc
\Zctdmalc{\ZcoosysfnA}

\Zcalmalc

The forward neutron calorimeter (FNC) \cite{nim:a354:479,nim:a394:121}
was installed in the HERA tunnel at 
\mbox{$\theta = 0$} degrees 
and at $Z = 106$ m from the interaction point in the 
proton-beam direction.
It is a lead-scintillator  calorimeter
which is segmented longitudinally into a front section, seven 
interaction lengths deep, and a rear section, three interaction lengths deep.
The front section is divided vertically into 14 towers,
allowing the separation of electromagnetic and hadronic showers
from the energy-weighted vertical width of the showers.  
The energy resolution for neutrons,
as measured in a beam test, 
is $\sigma(E_n)/E_n=0.65/\sqrt{E_n}$, with neutron energy, $E_n$,
 in GeV. 
Three planes of veto counters
 are used to reject events in which particles had
interacted with the inactive material
in front of the FNC. 
Magnet apertures limit the FNC acceptance to neutrons with 
production angles less than \mbox{$0.8$ mrad},
which corresponds to transverse momenta
$p_{\it T}\le E_n\theta_{\mbox{\rm\tiny max}}=0.74 x_L$~GeV.
The mean value of $p_{\it T}$ for the  data is 0.22 GeV. 





The luminosity was determined from the rate of the bremsstrahlung process
$ep \rightarrow e \gamma p$, where the photon was measured with a 
lead-scintillator calorimeter~\cite{desy-92-066,acpp:b32:2025} 
located at \hfill\\ \mbox{$Z = -107$ m}.

\section{Kinematics}

The kinematics of photoproduction at HERA
are specified by  the photon virtuality, $Q^2$, and the photon-proton 
center-of-mass energy,
$W$. The electron-proton center-of-mass energy,
$\sqrt{s}$, is related to $W$ by $W^2=ys$ where $y$ is the fraction of the
electron beam energy carried by the photon in the proton rest frame.

To describe the process 
$ep \rightarrow e^\prime \dstar nX$, four additional variables are used:
two for the neutron and two for the charmed meson.
They are:
\begin{itemize}
\item ($x_L$,$\theta_n$), the fractional energy and production angle of the produced neutron; 
only about half of the data have a $\theta_n$ measurement, therefore
all results discussed here are integrated over this variable
up to the maximum accepted angle of 0.8 mrad;
\item ($p_{\it T}$,$\eta$), the transverse momentum and pseudorapidity of the
produced $\dstar$ meson.
\end{itemize}

The measurement was performed in the following kinematic region:
$Q^2<1$ GeV$^2$, \mbox{$130<W<280$}~GeV, 
$|\eta(D^*)|<1.5$, $p_{\it T}(D^*)>1.9$ GeV, 
$x_L>0.2$ and $\theta_n<0.8$ mrad.
\section{Event Selection}
\label{sec-sel}


\subsection{Trigger}

A three-level trigger system was used to select events
online~\cite{zeus:1993:bluebook,uproc:chep:1992:222}. The selection
was based on energy deposits, tracking  and event timing.
The FNC was not used in the trigger.


\subsection{Photoproduction selection}
 
Photoproduction events were selected offline using cuts based
on the reconstructed primary vertex position, 
CAL energy deposits and the reconstructed
tracks of charged particles.
Events with a well-identified electron candidate 
in the CAL were removed. 
It was required that $\sum_i (E_i - p_{Z,i})>7$ GeV, where
the sum runs over all CAL cells and $p_{Z,i}$ is the $Z$ component of the
momentum vector assigned to each cell of energy $E_i$.
Tracking and CAL information
was combined to form energy flow objects (EFOs)  
\cite{epj:c1:81,briskin:phd:1998}. 
A cut was made on the Jacquet-Blondel~\cite{proc:epfacility:1979:391} estimator of $W^2$,
\mbox{$W^2_{\rm {JB}}=y_{\rm {JB}}s$}, where
\mbox{$y_{\rm {JB}} = \sum_i (E_i - E_{Z,i}) /2E_e$}, and
\mbox{$E_{Z,i} = E_i\cos\theta_i$};
$E_i$ is the energy of EFO
$i$ with polar angle $\theta_i$ with respect to the
measured $Z$-vertex of the event.
The sum runs over all EFOs. 
It was required that 
\mbox{$W_{\rm {JB}} < 265$ GeV}. These cuts
correspond to a true $W$ range of  \mbox{$130 < W < 280$ GeV} and $Q^2 < 1$ ~GeV$^{2}$
with the median  $Q^2 \approx 10^{-3}$~GeV$^2$.

\subsection{$\mathbf{\dstar (2010)}$ reconstruction}

The inclusive charm sample was selected by identifying events
containing a charmed meson.
The $\dstar$ selection cuts are based on the decay channel: 
\mbox{$D^{* +} \rightarrow (D^0 \rightarrow K^- \pi^+) \pi^+_s$}
(+~charge conjugate),
where $\pi_s$ indicates the ``slow'' pion~\cite{prl:35:1672}. 
 Only tracks assigned to the primary event vertex and
with hits in at least three superlayers of the CTD were considered.
The combinatorial background was reduced and the 
kinematic phase space defined by requiring:
     the transverse momentum of the kaon and pion candidates to satisfy
  \mbox{$p_{\it T}(K) > 0.45$ GeV}, 
      \mbox{$p_{\it T}(\pi) > 0.45$ GeV} and  
      \mbox{$p_{\it T}(\pi_s)> 0.12$ GeV};
the transverse momentum of the $\dstar$ to be greater than 1.9 GeV
 and the pseudorapidity of the $\dstar$ to satisfy $|\eta(\dstar)|<1.5$.

Since no particle identification was performed, the $K$ and $\pi$ masses were 
alternately attributed to the decay products of the candidate $D^0$ meson.  
Those $D^0$ candidates that had an invariant mass between 
1.80 and 1.92 GeV were required to have a
mass difference 
$\Delta M = M(K \pi \pi_s) - M(K \pi)$  between 0.1435 and 0.1475 GeV. 
The combinatorial background  was estimated from
the mass-difference distribution for wrong-charge combinations,
in which both tracks forming the $D^0$ candidates have the same charge
and the third track has the opposite charge.

\subsection{Neutron reconstruction}

Events with a leading neutron were selected from the inclusive charm sample
by requiring a large energy deposit ($E_n>184$ GeV) in the FNC.
 Protons, photons, and neutrons are separated by their position in the
detector, as well as by the shower width. Scattered protons are 
deflected
by the HERA  magnets and strike the top part of the FNC.
Photons can be identified and removed from the sample because the
transverse spread of electromagnetic showers is
much less than that of hadronic showers.


Events with particles that started to shower before reaching the FNC were
removed by requiring that the scintillator veto counter  had an energy 
deposit below that of a minimum-ionizing particle.
Events with wide showers,  
inconsistent with originating from a single high-energy hadron, were removed.


\subsection{Final event sample}



\begin{figure}[htbp!]
\centerline{\epsffile{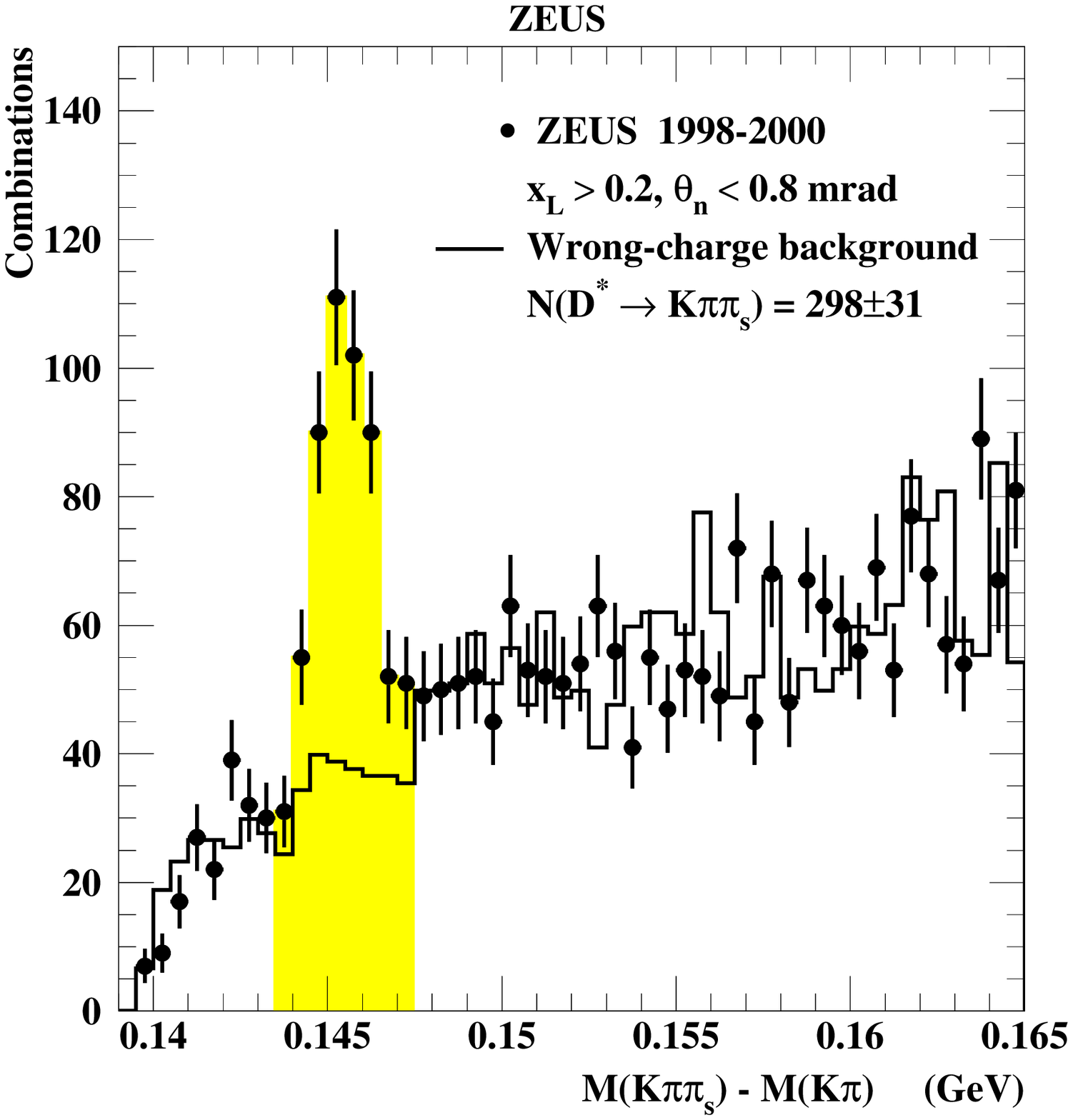}}
\vspace{-5.9cm} \hspace{10.2cm} \epsfxsize=5.cm\epsfysize=3.5cm \epsffile{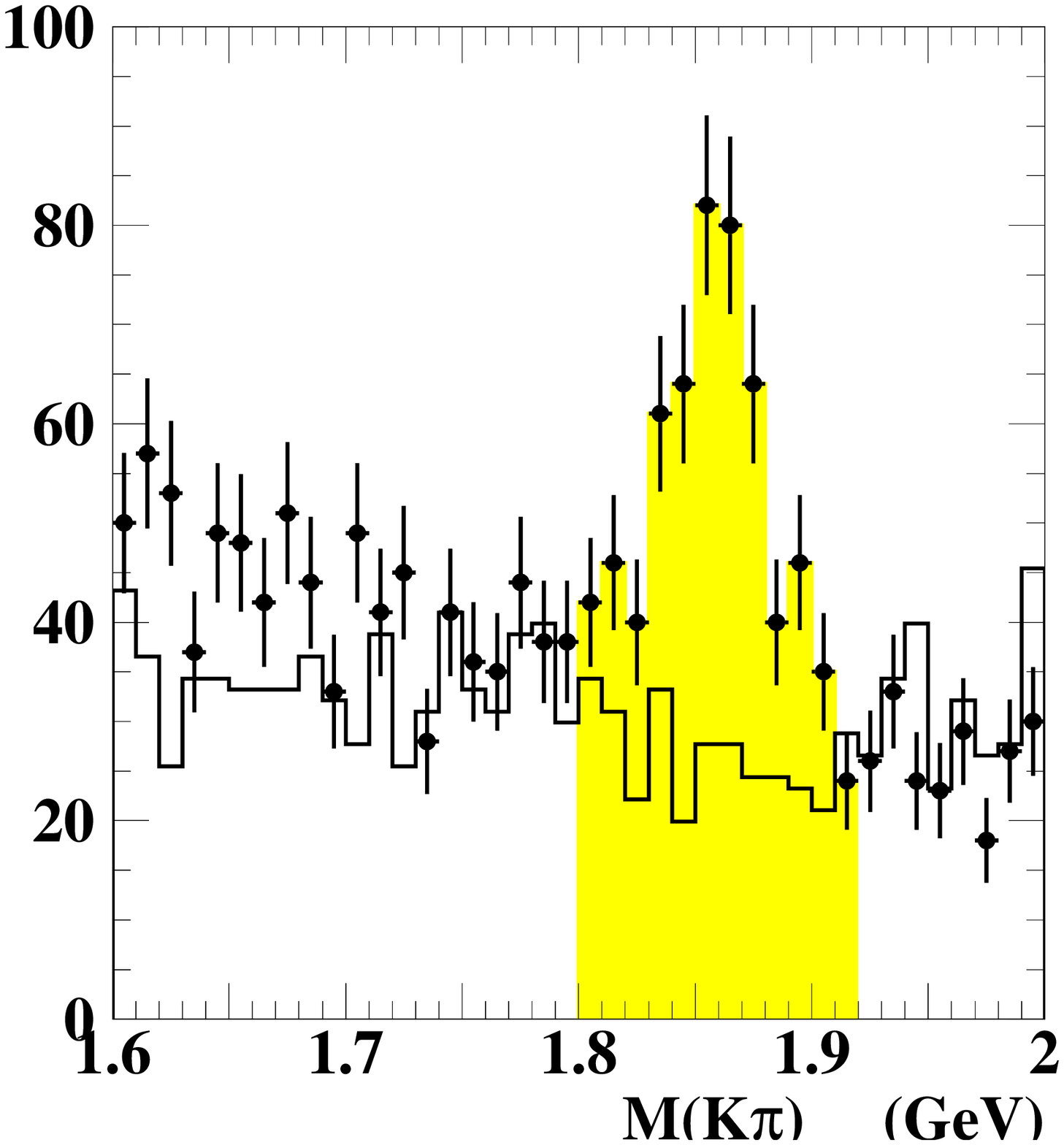}
\vspace{2.3cm} 
\caption{
The data points show the neutron-tagged $\Delta M$ distribution 
for right-charge track combinations. The solid line shows the
wrong-charge combinations normalized to the right-charge
combinations in the region \mbox{$0.15<\Delta M<0.165$} GeV
outside the $\dstar$ mass window 
$0.1435<\Delta M<0.1475$ GeV which is shown shaded. 
The signal observed in the $M(K \pi)$ distribution
for events within the mass window $0.1435 < \Delta M < 0.1475$ GeV is 
shown as an inset. The solid line also shows the
wrong-charge combinations normalized to the right-charge combinations as
before.}
\label{fig-fig1}
\end{figure}

The $\Delta M$ distribution for the neutron-tagged sample is
shown in Fig.~\ref{fig-fig1}, along with the wrong-charge combinations. 
A  prominent $\dstar$ signal is observed.  
The signal observed in the $M(K \pi)$ distribution
for events within the mass window $0.1435 < \Delta M < 0.1475$ GeV is 
shown as an inset.

After the wrong-charge background subtraction, 298 $\pm$ 31 $\dstar$ 
mesons were found. The same background subtraction  procedure, applied to the
inclusive $\dstar$ sample, gave \mbox{14\,743 $\pm$ 253 events}. 


\section{Monte Carlo simulation and acceptance corrections}
\label{sec-mc}

A {\sc Geant}-based \cite{tech:cern-dd-ee-84-1} Monte Carlo (MC) 
simulation was used to calculate selection efficiencies and 
correction factors for the charmed meson.
Three different 
event generators were used:  {\sc Rapgap}~2.08/06\cite{cpc:86:147}
for evaluating the nominal corrections, 
 {\sc Herwig}~6.301\cite{cpc:67:465} 
 and {\sc Pythia}~6.156\cite{cpc:135:238} as 
systematic checks. 
{\sc Rapgap} and {\sc Pythia} use the Lund string model
for  hadronization. {\sc Herwig} uses a cluster model.
The events generated with {\sc Rapgap} for acceptance calculations
were produced using OPE 
for the production of the leading  neutron,
with  the pion flux factor from Eqs. (\ref{eq:flux}) and (\ref{flu2})
and  the GRV parametrization \cite{grv} for the pion PDF. 
Inclusive  {\sc Rapgap}, employing the Lund string model
instead of OPE for neutron production, was produced for
comparisons to the final measurements. 
The leading neutron is also produced  via the
 Lund string model in {\sc Pythia}. In {\sc Herwig}, it is
produced via the cluster model. 
The proton PDFs parametrizations used were CTEQ5L \cite{cteq5l} for
{\sc Pythia} and  {\sc Herwig},  and CTEQ4D \cite{cteq4d} for 
inclusive {\sc Rapgap}.
The photon PDF GRV-G LO \cite{grv-g} was used  in {\sc Pythia} and  
{\sc Herwig}, and GRS LO \cite{grs} in {\sc Rapgap}.
The mass of the charm quark was set to 1.5 GeV.
The fraction of $c$ quarks hadronizing to 
a $D^*$ meson was set to $f(c\rightarrow D^*)=0.235$ \cite{leonid}.
Both direct and resolved photon processes 
for charm production were generated,
in proportion to their predicted cross sections.
Effects of neutron rescattering  were not taken into account in
the simulation.

For all the MC samples used to evaluate the acceptances, events with at least one $\dstar$ 
decaying in the appropriate decay channel were selected and passed through the
 ZEUS detector and trigger simulations as well as the event-reconstruction package.

Since the  $D^*$ and neutron were independently detected and their kinematics
 largely uncorrelated, the acceptances
for the two particles factorize.
The selection efficiencies and correction factors for the
neutron 
calculated 
for previous analyses \cite{np:b637:3} were used. 
The overall acceptance of the FNC, which includes the beam-line geometry 
 and the angular distribution of the neutrons,
is about 25\% 
for neutrons with \mbox{$x_L>0.2$}
and \mbox{$\theta_n<0.8$ mrad}. 

The differential cross section for $\dstar$ photoproduction associated with a 
leading neutron was evaluated in terms of a given variable  $Y$ as
$d\sigma/dY= N/(A_{\rm FNC}\cdot A_{D^*}\cdot B \cdot \Delta Y$)
where N is the number of $D^*$ found in the final sample in a bin of
size $\Delta Y$, $A_{\rm FNC}$ is the acceptance for the neutron detection
in the FNC, $A_{D^*}$ is the acceptance for the $D^*$ reconstruction 
and $B$ is the branching ratio for the selected decay mode. A value
of $B =2.57 \%$ \cite{pr:d66:010001} was used.

\section{Systematic uncertainties}
\label{sec-sys}

For the $\dstar$ measurement the major sources of systematic uncertainty
are listed below (the relative uncertainty on 
$\dstar$ acceptance is shown in parentheses):


\begin{itemize}
\item the selection of photoproduction events and $\dstar$ candidates. 
       Variations were made in the
      $W_{\rm {JB}}$ (${+4.5}$ \%) and  $Z$-vertex ($^{+1.5}_{-0.3}$ \%) cuts;

\item the $p_{\it T}$ of the pion and kaon cuts. These were
         varied according to their
         resolutions ($^{+1.2}_{-1.7}$ \%);
\item variation of the mass windows.
      The $\Delta M$ window used for the extraction of the
      $\dstar$ was widened symmetrically by 0.5~MeV.
      The $M(D^0)$ window was widened and reduced symmetrically
      by 5 MeV ($^{+3.4}_{-3.6}$~\%);
\item the $\Delta M$ region for the normalization of the 
      wrong-charged  combinations. This was changed from 
      \mbox{$0.15 - 0.165$ GeV} to 
      \mbox{$0.15 - 0.163$ GeV}  (+2.4 \%);
\item the MC model dependence.
      {\sc Herwig} ($-$4.8 \%) and {\sc Pythia} (+2 \%) were used instead of {\sc Rapgap};
\item the fraction of resolved photon events in the MC 
        was lowered by 20\% and 
        raised by 10 \% ($^{+0.8}_{-1.8}$ \%);
\item  the CAL energy scale. This was varied within its uncertainty of $\pm 3\%$ ($\pm 0.9$ \%).

\end{itemize} 

An extensive discussion of the systematic effects related to  
the neutron measurement is given elsewhere \cite{np:b637:3}.
Here, the major sources of systematic  uncertainty and their effect 
on the FNC acceptance (shown in parentheses) is listed: 
\begin{itemize}
\item the uncertainty in the angular distribution of the
      neutrons ($\pm 4$\% for $x_L<0.82$, $\pm 7$\%~for~$x_L>0.82$); 
\item the uncertainty in the overall FNC energy scale of $\pm 2$\%
        (less than 4\% effect for \mbox{$x_L<0.82$}, 
        $^{+14}_{-16}$\%~for $x_L>0.82$ );
\item the normalization uncertainty arising from
      proton-beam-gas interactions overlapping with 
      photoproduction events, 
      the uncertainty in the amount of dead material in the 
      beam line, and the uncertainties from the veto cuts ($\pm 5$\%).  
\end{itemize}

All above errors were added in quadrature separately for the
positive and negative variations
to determine the overall systematic uncertainty.
The overall normalization has additional uncertainties of 2.2\% due 
to the luminosity  measurement and  2.5\%  due to 
the knowledge of branching ratios.

Sources of systematic uncertainty in the ratio measurement
were studied in a similar manner to those for the cross-section measurements.
There is a cancellation between the common systematic uncertainties
originating from the selection of inclusive photoproduction events,
the selection of $\dstar$ candidates and the background estimation.
The remaining contributions are  those from the model dependence of
the acceptance corrections used in the evaluation of the inclusive
$D^{*\pm}$ photoproduction cross sections and from the neutron measurement
uncertainties. 

\section{Results}
\label{sec-results}


The integrated cross section  for the reaction $ep \rightarrow e^\prime \dstar nX$
in the kinematic region
\mbox{$Q^2<1$ GeV$^2$},
\mbox{$130<W<280$ GeV},
\mbox{$|\eta(D^*)|<1.5$},
\mbox{$p_{\it T}(D^*)>1.9$ GeV}, 
$x_L>0.2$ ~and $\theta_n<0.8$ mrad is 
 \[
 2.08\pm 0.22({\rm stat.})^{+0.12}_{-0.18}({\rm syst.})\pm0.05({\rm B.R.})\ {\rm nb,}
 \]
where the final uncertainty arises from the uncertainty of the branching
ratios for the $D^*$ and $D^0$. The luminosity uncertainty was included 
in the systematic uncertainty. 
The predicted cross sections from the models are $3.0$~nb for 
{\sc Herwig}, $4.6$~nb for  {\sc Pythia}, $2.6$~nb for inclusive
{\sc Rapgap} and $2.0$~nb for {\sc Rapgap} with OPE. 

Table 1 and Fig.~\ref{fig-fig3} show the differential 
cross sections for neutron-tagged $\dstar$ production
as a function of $W$, $p_{\it T} (D^*)$ and $\eta (D^*)$. The differential 
cross section as a function of  $x_L$ is shown in
Table 1 and Fig.~\ref{fig-fig4}.
 
The inclusive $\dstar$ cross section
was  measured and found to agree with previous 
measurements~\cite{epj:c6:67} in a similar kinematic range.
Table~\ref{tab-rat}  and Fig.~\ref{fig-fig5}  show the measured 
ratios of neutron-tagged to inclusive $\dstar$ production as a function of different
 kinematic variables.
Over the whole measured kinematic range the ratio 
 is
\[ 
r^{D^*} = 8.85\pm 0.93({\rm stat.})^{+0.48}_{-0.61}({\rm syst.})\ \%
\]
which is shown superposed on Fig.~\ref{fig-fig5}. 
 The $\chi^2$
per degree of freedom with respect to the overall ratio
are 0.27, 1.65 and 0.09
for the $W$, $p_{\it T}(D^*)$ and $\eta(D^*)$ distributions. 
Within the experimental
uncertainties neutron-tagged $\dstar$ production is compatible with 
being a constant fraction of inclusive $\dstar$ production, independent
of  the $\dstar$ kinematics.

\section{Discussion}

\begin{figure}[htbp!]
\centerline{\epsffile{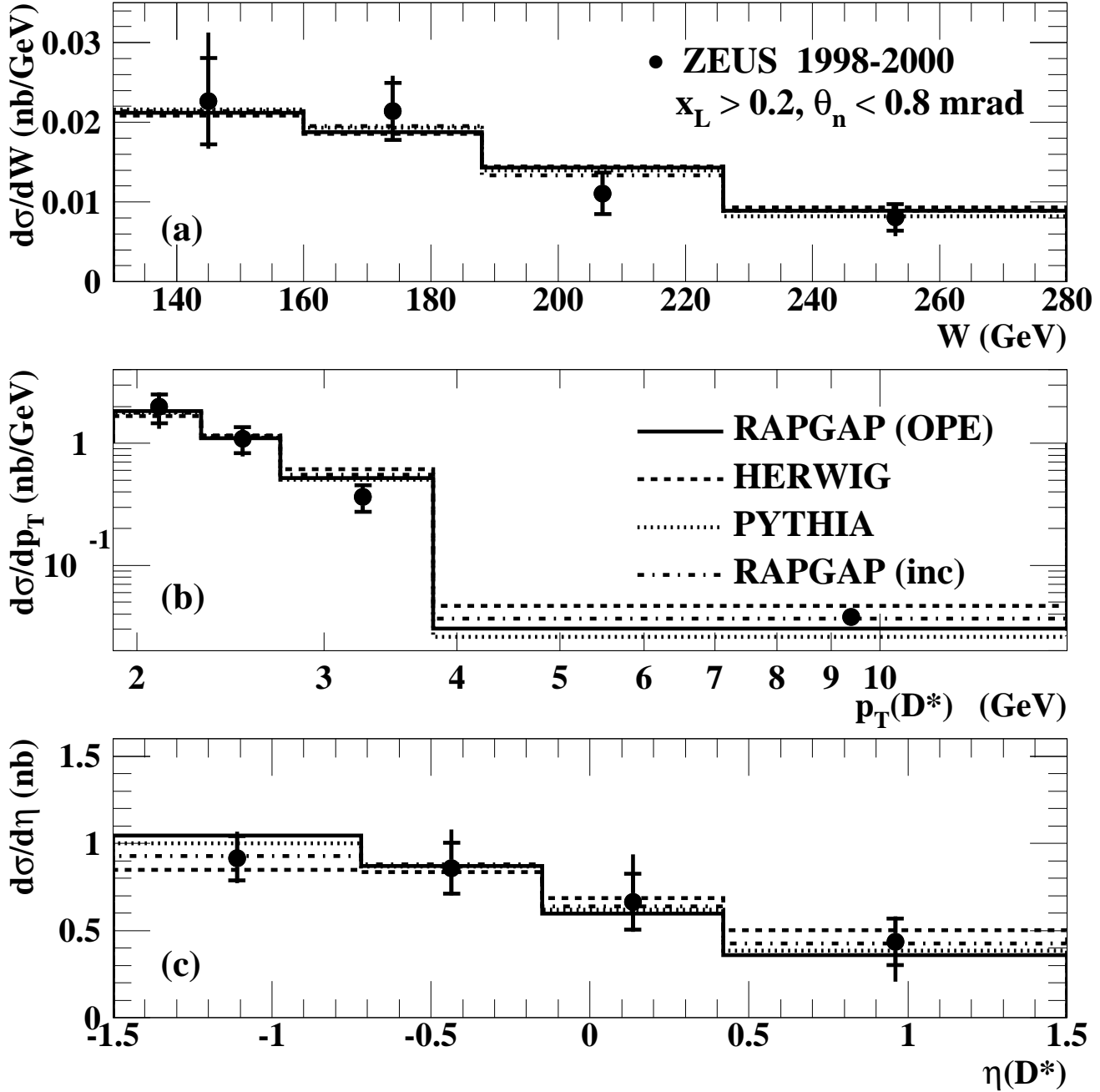}}
\caption{
The points show the differential cross sections 
for neutron-tagged $\dstar$ production
as a function of $W$, $p_{\rm T} (D^*)$ and $\eta (D^*)$
for $x_L>0.2$ and  $\theta_n<0.8$ mrad.
The error bars displayed on the plots denote the statistical
uncertainty (inner) and the quadratic sum of the statistical and 
the systematic uncertainties (outer). The uncertainties due 
to the luminosity measurement 
and  the branching ratios  
are not shown. 
The predictions of Monte Carlo models 
normalized to the data are also shown.
}
\label{fig-fig3}
\end{figure}

\begin{figure}[htbp!]
\centerline{\epsffile{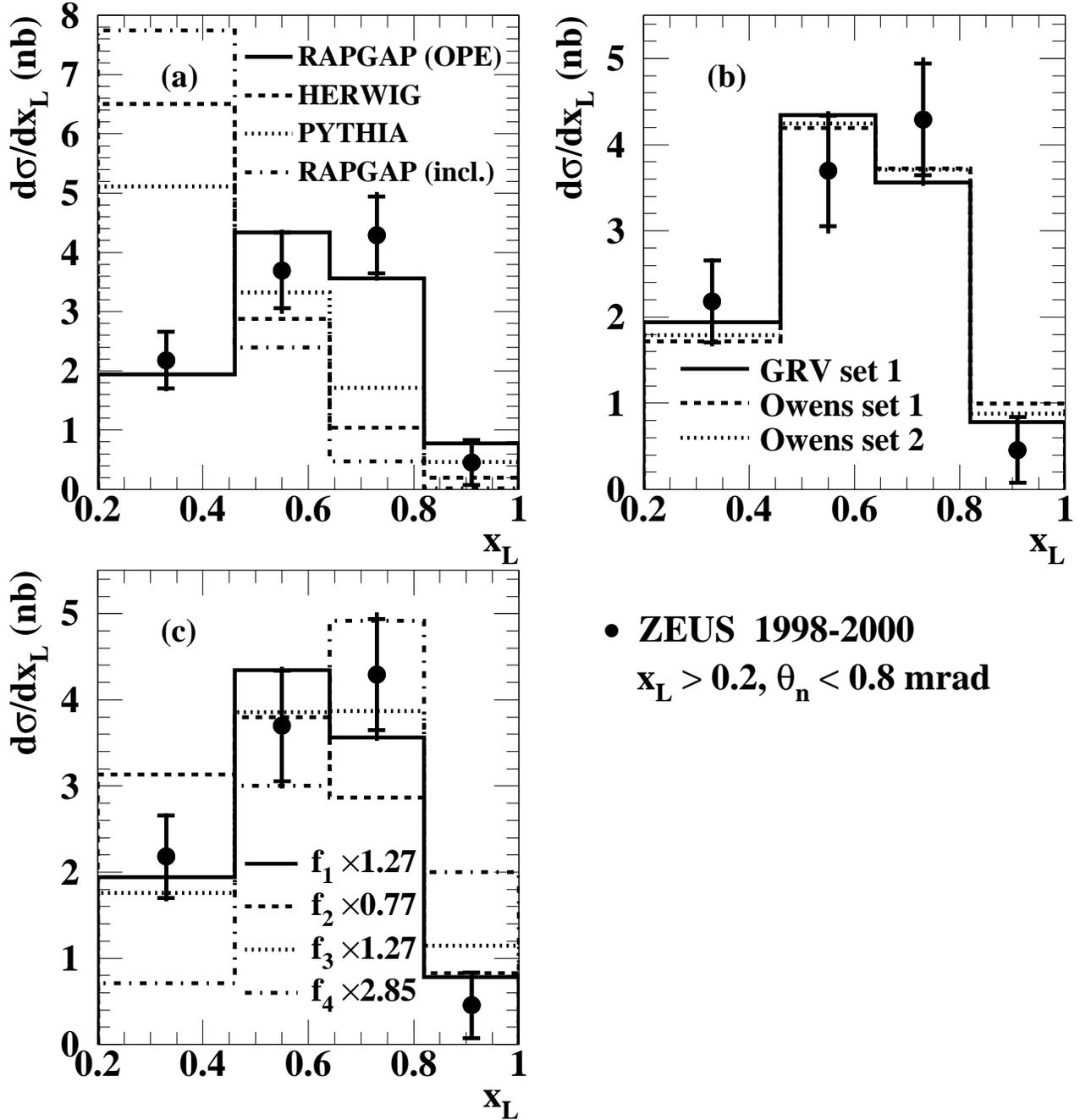}}
\vspace{-0.5cm}
\caption{\protect The points show the differential cross sections 
for neutron-tagged $\dstar$ production
as a function of $x_L$.
The histograms show 
the predictions of Monte Carlo models (a) {\sc Rapgap} with OPE 
(solid histogram), {\sc Herwig} (dashed), 
{\sc Pythia} (dotted), and inclusive {\sc Rapgap}
(dashed-dotted); (b) {\sc Rapgap} with OPE and pion PDF
parametrizations from
 GRV set 1 (solid), Owens sets 1 (dashed) and 2  (dotted); 
(c) {\sc Rapgap} with OPE and flux factors $f_1$-$f_4$ from 
Eqs.~(\ref{flu2})-(\ref{flu3}). The error bars displayed on the plots denote
 the statistical
uncertainty (inner) and the quadratic sum of the statistical and 
the systematic uncertainties (outer). The uncertainties due 
to the luminosity measurement 
and  the branching ratios  
are not shown. 
All distributions are normalized to the data. The numbers in (c) are
the normalization factors.}
\label{fig-fig4}
\end{figure}

\begin{figure}[htbp!]
\centerline{\epsffile{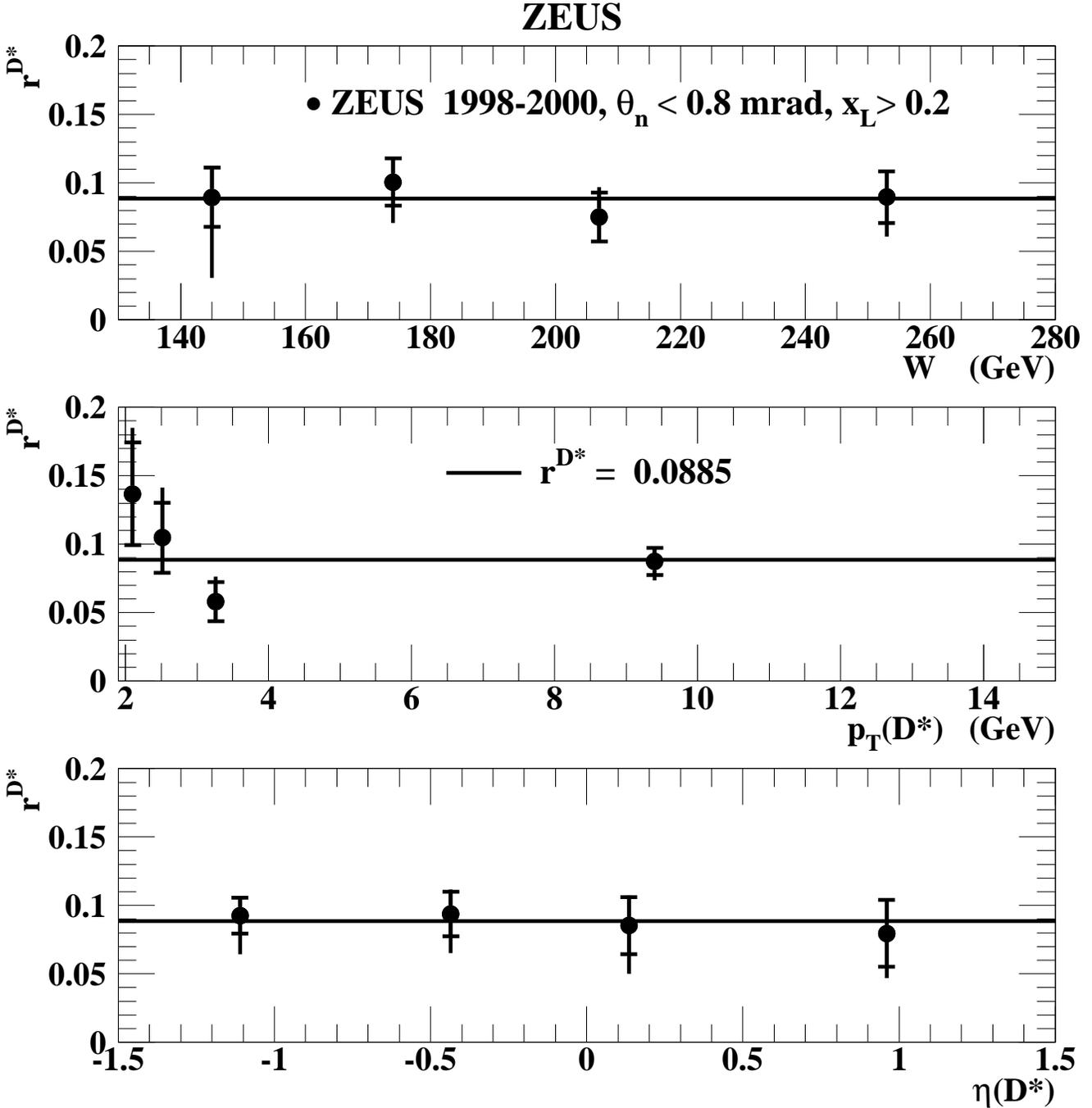}}
\caption{The ratio of neutron-tagged $\dstar$ production
to inclusive $\dstar$ production as a function of
$W$, $p_{\rm T} (D^*)$ and $\eta (D^*)$ for
$x_L>0.2$ and $\theta_n<0.8$ mrad.
The error bars displayed on the plots denote the statistical
uncertainty (inner) and the quadratic sum of the statistical and 
the systematic uncertainties (outer). 
The line superposed on the figures shows the overall
ratio of neutron-tagged $\dstar$ to inclusive
$\dstar$ events.
}
\label{fig-fig5}
\end{figure}

The experimental results in Figs.~\ref{fig-fig3} and \ref{fig-fig4}a  
are compared  to the predictions, normalized to the data, of 
the MC models {\sc Rapgap}  with OPE for
leading-neutron production, {\sc Herwig}, {\sc Pythia} 
and inclusive {\sc Rapgap}.  
In Fig.~\ref{fig-fig3}, the  
agreement in shape between data and MC models is satisfactory in all cases. 
 Fig.~\ref{fig-fig4}a, however, shows that only {\sc Rapgap} with OPE
 agrees with the measured
neutron energy distribution  seen in the data.
 A similar result was obtained in the study of neutron-tagged 
dijet photoproduction\cite{np:b596:3}.

In principle the  data allow the pion PDF to be probed in the range of the
parton fractional momenta
$10^{-3}<x_\pi<10^{-2}$.
Figure \ref{fig-fig4}b shows the differential cross section in $x_L$
compared to the predicted cross sections from
{\sc Rapgap} with OPE, based on different parametrizations for 
the pion structure 
function: GRV set 1 and Owens sets 1 and 2.
The data have little sensitivity, as in the case of neutron-tagged 
dijet photoproduction\cite{np:b596:3}, to the choice of
the pion structure function. Even with an extreme choice of a pion 
structure function, e.g.  a completely flat gluon distribution, or 
a parametrization identical to the proton structure function, little
variation is seen in the
predictions. 

Figure \ref{fig-fig4}c shows the data compared to the predictions
 of {\sc Rapgap} with OPE for the four flux factors discussed 
in Section \ref{sec:opeint}. All  {\sc Rapgap}
distributions are normalized to the data and the resulting normalization factors
are given in the figure. 
Fluxes $f_1$ and $f_3$ give 
similar results, being compatible with the 
data both in shape and normalization.
 The fluxes $f_2$ and  $f_4$ are disfavored by the 
shape of the data, and the 
latter predicts cross sections almost three times smaller than data.

The independence of $r^{D^*}$ on $W$, $p_{\it T} (D^*)$ and $\eta (D^*)$,
shown in Fig.~\ref{fig-fig5}, supports the hypothesis of
vertex factorization, in agreement 
with  previous studies of leading-neutron production 
\cite{pl:b384:388,np:b596:3,np:b637:3}. The predicted ratios
 from the models are: 0.24 for 
{\sc Herwig}, 0.29 for  {\sc Pythia}, 
and 0.18 for {\sc Rapgap}. 

The  ratio for charm production, ~$r^{D^*}$, is in agreement 
with the analogous ratio previously measured for neutron-tagged DIS,
\mbox{$Q^2>4$ GeV$^2$}  \cite{np:b637:3},
\[
r^{\rm DIS} = 8.0\pm 0.5\ \%,
\]
but lies above the ratio previously measured for neutron-tagged  
inclusive photoproduction at \mbox{$W=207$}~GeV \cite{np:b637:3},
\[
 r^{\gamma p} = 5.7\pm 0.4\ \%,
\] 
as expected from the rescattering effects within the OPE model
 (see Section \ref{sec:resc}). The errors quoted are the quadratic sum of 
the statistical and systematic uncertainties. 

For photoproduced dijets
the neutron-tagged to inclusive ratio
has only been measured for \mbox{$x_L>0.49$}. 
In this kinematic region, the measured $\dstar$ ratio 
and their corresponding ratios previously measured for DIS, 
inclusive photoproduction 
 \cite{np:b637:3}, and  photoproduction 
of dijets with transverse energy \mbox{$E_T^{\rm jet}>6$ GeV}  \cite{np:b596:3} are:
\begin{eqnarray}
r^{D^*}(x_L>0.49) &=& 6.55\pm 0.76({\rm stat.})^{+0.35}_{-0.45}({\rm syst.}) \ \%\nonumber \\
r^{\rm DIS}(x_L>0.49) &=& 5.8\pm 0.3 \ \%\nonumber \\
r^{\rm jj}(x_L>0.49) &=& 4.9\pm 0.4 \ \%\nonumber \\
r^{\gamma p}(x_L>0.49) &=& 4.3\pm 0.3 \ \%\nonumber.
\end{eqnarray}
The results are compatible with the rescattering 
hypothesis  described in Section 2.3.

\section{Summary}

The photoproduction of $\dstar$ mesons associated with a leading
neutron has been studied in $ep$ interactions at HERA
in the kinematic region
\mbox{$Q^2<1$ GeV$^2$},
\mbox{$130<W<280$ GeV},
\mbox{$|\eta (D^*)|<1.5$},
\mbox{$p_{\it T} (D^*)>1.9$ GeV}, 
\mbox{$\theta_n<0.8$ mrad} 
and \mbox{$x_L>0.2$}.
The Monte Carlo models {\sc Rapgap}, {\sc Herwig} and {\sc Pythia}
give a satisfactory
description of the $\dstar$ kinematics, but only {\sc Rapgap} with
one-pion exchange satisfactorily describes the  leading-neutron energy distribution. 
The results show sensitivity to the choice
of pion flux factor.
The ratio of neutron-tagged $\dstar$ photoproduction to 
inclusive $\dstar$ photoproduction is 
\mbox{$r^{D^*}=8.85\pm 0.93({\rm stat.})^{+0.48}_{-0.61}({\rm syst.})$\%}.

The ratio of neutron-tagged $\dstar$ photoproduction to
inclusive $\dstar$ photoproduction is 
constant
as a function of  $W$, $p_{\it T} (D^*)$ and $\eta (D^*)$, in agreement with
the hypothesis of vertex
factorization. 
This ratio is consistent with the analogous ratio
in deep inelastic scattering, 
but both are about 30\% higher than the corresponding ratio
for inclusive photoproduction, suggesting that the presence of a 
hard scale 
enhances the fraction of events with a leading neutron 
in the final state.
 
\section*{Acknowledgements}

We thank the DESY Directorate for their strong support and encouragement, and 
the HERA machine group for their diligent efforts.  We are grateful for the 
support of the DESY computing and network services.  The design, construction
and installation of the ZEUS detector have been made possible owing to the 
ingenuity and effort of many people 
who are not listed as authors.  


{
\def\bibname{\Large\bf References}
\def\refname{\Large\bf References}
\pagestyle{plain}
\ifzeusbst
  \bibliographystyle{./BiBTeX/bst/l4z_default}
\fi
\ifzdrftbst
  \bibliographystyle{./BiBTeX/bst/l4z_draft}
\fi
\ifzbstepj
  \bibliographystyle{./BiBTeX/bst/l4z_epj}
\fi
\ifzbstnp
  \bibliographystyle{./BiBTeX/bst/l4z_np}
\fi
\ifzbstpl
  \bibliographystyle{./BiBTeX/bst/l4z_pl}
\fi
{\raggedright
\bibliography{./BiBTeX/user/syn.bib,%
	      ./BiBTeX/user/extra.bib,%
              ./BiBTeX/bib/l4z_articles.bib,%
              ./BiBTeX/bib/l4z_books.bib,%
              ./BiBTeX/bib/l4z_conferences.bib,%
              ./BiBTeX/bib/l4z_h1.bib,%
              ./BiBTeX/bib/l4z_misc.bib,%
              ./BiBTeX/bib/l4z_old.bib,%
              ./BiBTeX/bib/l4z_preprints.bib,%
              ./BiBTeX/bib/l4z_replaced.bib,%
              ./BiBTeX/bib/l4z_temporary.bib,%
              ./BiBTeX/bib/l4z_zeus.bib}}
}
\vfill\eject

%
\begin{table}[hbt]
\begin{center}
\begin{tabular}{|l|c|c|} \hline
%
$W$ Range ($\gev$)  & $\diff \sigma / \diff W \pm ({\rm stat.})\pm ({\rm syst.})$ (nb/GeV) \\ \hline
$130$ - $160$  & $0.0226 \pm 0.0054 _{-0.0025 }^{+0.0067 } $ \\
$160$ - $188$  & $0.0214 \pm  0.0036_{-0.0004 }^{+0.0027 } $ \\
$188$ - $226$  & $0.0111 \pm  0.0026_{-0.0007 }^{+0.0008 } $ \\
$226$ - $280$  & $0.0080 \pm  0.0017_{-0.0017 }^{+0.0007 } $ \\ \hline
%
$p_{\it T}(\dstar)$ Range ($\gev$)  & $\diff \sigma / \diff p_{\it T} (\dstar) \pm ({\rm stat.})\pm ({\rm syst.})$ (nb/GeV) \\ \hline
$1.9$ - $2.3$    & $ 1.991  \pm 0.531 _{-0.421 }^{+0.317 } $ \\
$2.3$ - $2.73$  & $  1.089 \pm 0.261  _{-0.168 }^{+ 0.144 } $ \\
$2.73$ - $3.8$ &  $ 0.364 \pm 0.090 _{-0.018  }^{+0.055 } $ \\
$3.8$ - $15$   &  $ 0.038 \pm  0.004_{-0.004 }^{+0.003 } $ \\ \hline%
$\eta(\dstar)$ Range  & $\diff \sigma / \diff \eta (\dstar) \pm ({\rm stat.})\pm ({\rm syst.})$ (nb) \\ \hline
$(-1.5\ )$ - $(-0.72)$   & $ 0.914 \pm 0.126 _{-0.067 }^{+0.087 } $ \\
$(-0.72)$ - $(-0.15)$    & $ 0.858 \pm 0.147 _{-0.055 }^{+0.164 } $ \\
$(-0.15)$ - $(+0.42)$    & $ 0.665 \pm 0.161 _{-0.064 }^{+0.222 } $ \\
$(+0.42)$ - $(+1.5\ )$   & $ 0.436 \pm 0.132 _{-0.186 }^{+0.061 } $ \\ \hline
%
$x_L$ Range &  $\diff \sigma / \diff x_L \pm ({\rm stat.})\pm ({\rm syst.})$ (nb) \\ \hline
$0.2$  - $0.46$  & $  2.18\pm 0.48 _{-0.20 }^{+0.15 } $ \\
$0.46$ - $0.64$  &  $  3.70\pm 0.64 _{-0.34 }^{+0.25 } $ \\
$0.64$ - $0.82$  &  $  4.29\pm 0.65  _{-0.42 }^{+0.32 } $ \\
$0.82$ - $1.0$   &  $  0.456\pm 0.381 _{-0.086 }^{+0.073 } $ \\ \hline
\end{tabular}
\caption{Values of the differential cross sections 
for neutron-tagged $\dstar$ photoproduction
(\mbox{$Q^2<1$ GeV$^2$} and \mbox{$\theta_n<$ 0.8 mrad})
with respect to 
$W$, $p_{\it T}(\dstar)$, $\eta(\dstar)$ and $x_L$.
}
\label{tab-xsd}
\end{center}
\end{table}

\begin{table}[hbt]
\begin{center}
\begin{tabular}{|l|c|} \hline
$W$ Range ($\gev$) & $ r^{D^*} \pm ({\rm stat.})\pm ({\rm syst.})$  \\ \hline
$130$ - $160$    & $0.089  \pm 0.022 _{- 0.055 }^{+ 0.004 } $ \\
$160$ - $188$    & $ 0.101  \pm  0.017 _{- 0.025 }^{+0.006  } $ \\
$188$ - $226$    & $ 0.075  \pm  0.018 _{-0.004  }^{+0.012  } $ \\
$226$ - $280$    & $ 0.090  \pm  0.019_{- 0.022 }^{+0.006  } $ \\ \hline
$p_{\it T}(\dstar)$ Range ($\gev$) & $r^{D^*}  \pm ({\rm stat.})\pm ({\rm syst.})$  \\ \hline
$1.9$ - $2.3$  & $  0.137 \pm 0.038  _{- 0.007 }^{+ 0.030 } $ \\
$2.3$ - $2.73$ & $  0.105   \pm 0.026   _{- 0.005 }^{+ 0.026  } $ \\
$2.73$ - $3.8$ & $ 0.058   \pm 0.014  _{- 0.003 }^{+ 0.011 } $ \\
$3.8$ - $15$   & $  0.087 \pm 0.010  _{-0.009 }^{+0.004  } $ \\ \hline
$\eta(\dstar)$ Range & $r^{D^*}  \pm ({\rm stat.})\pm ({\rm syst.})$  \\ \hline
$(-1.5\ )$ - $(-0.72)$  & $ 0.093 \pm 0.013 _{-0.025 }^{+0.005 } $ \\
$(-0.72)$ - $(-0.15)$   & $ 0.097 \pm 0.016  _{-0.024 }^{+0.007 } $ \\
$(-0.15)$ - $(+0.42)$   & $ 0.085 \pm 0.021  _{-0.028 }^{+0.007} $ \\
$(+0.42)$ - $(+1.5\ )$  & $ 0.080 \pm 0.024  _{-0.022 }^{+0.004 } $ \\ \hline
\end{tabular}
\caption{Values of the ratios of the differential cross sections 
for neutron-tagged to inclusive $\dstar$  photoproduction
(\mbox{$Q^2<1$ GeV$^2$}, $x_L>0.2$ and \mbox{$\theta_n<$ 0.8 mrad})
with respect to 
$W$, $p_{\it T}(\dstar)$ and $\eta(\dstar)$.
}
\label{tab-rat}
\end{center}
\end{table}



\clearpage

\clearpage

\clearpage

%
%
\end{document}